\documentstyle[twoside,fleqn,espcrc2]{article}

\newcommand{\AmS}{{\protect\the\textfont2
  A\kern-.1667em\lower.5ex\hbox{M}\kern-.125emS}}

\title{Rare Kaon Decays}

\author{Gerhard Buchalla\address{Stanford Linear Accelerator Center, \\ 
        Stanford University, Stanford, CA 94309, U.S.A.}%
        \thanks{Work supported by the Department of Energy under contract
                 DE-AC03-76SF00515.}}

\begin{document}

\begin{abstract}
The phenomenology of rare kaon decays is reviewed, with emphasis
on tests of standard model flavordynamics.
\end{abstract}

\maketitle

\section{INTRODUCTION}

The physics of kaons has played a major role in the development
of particle physics. The concept of strangeness, with its implications
for the quark model, the discovery of P and CP violation and the
anticipation of charm and the GIM mechanism have all emerged from the
study of K mesons. Today, rare decays of kaons continue to be an active
field of investigation. For several topics of current interest
kaon physics holds the promise of providing important insights:
\begin{itemize}
\item
Rare kaon decays probe the flavordynamics of the standard model (SM),
i.e. the physics of quark masses and mixing. This part of the theory
is related to electroweak symmetry breaking, the sector of the standard
model that is least understood and contains most of the free model
parameters.
\item
The sensitivity of rare kaon phenomena, like $K-\bar K$ mixing, to
energies  higher than the kaon mass scale proved to be a very useful
source of information on the charm quark, even before it was discovered
\cite{GL}. In quite the same spirit rare decays of K mesons allow to
probe the physics of top quarks. Many of them are strongly sensitive
to the top quark mass and can yield results on the CKM couplings
$V_{td}$ and $V_{ts}$, information hardly accessible directly in
top quark decays.
\item
Among the most important current problems in high energy physics is
the poorly understood topic of CP violation. All of the, very few,
experimental facts known to date about this fundamental asymmetry 
derive from a handful of $K_L$ decay modes ($K_L\to\pi\pi$,
$\pi l\nu$, $\pi^+\pi^-\gamma$) and can so far be all accomodated
by a single complex parameter $\varepsilon_K$. More detailed insight
into CP violation will be possible, for instance, by 
the precision studies offered through measuring theoretically clean
rare K decays, such as $K_L\to\pi^0\nu\bar\nu$.
\item
Beyond rare processes that are strongly suppressed in the standard
model, one may search for decays that are entirely forbidden within
this framework and therefore particularly clear signals of new physics.
Promising examples are lepton-family-number violating modes like
$K_L\to\mu e$. Experiments are planned to probe the corresponding
branching ratios down to a level of $\sim 10^{-12}$. This accuracy
may be translated into a sensitivity to scales of new physics of
a few hundred $TeV$, although precise values are model dependent.
Sensitivity to such high energy scales seems very difficult to attain
by any other method.
\item
Some rare and radiative decay modes of K mesons (such as
$K^+\to\pi^+l^+l^-$ or $K_L\to\pi^0\gamma\gamma$) are dominated by
long-distance hadronic physics and therefore less useful for the
study of short-distance flavordynamics. Still these cases are of
considerable interest to test the low energy structure of QCD as 
described within the framework of chiral perturbation theory.
\end{itemize}

It is evident that the area of rare kaon decays covers a rather wide
range of important topics. In the following talk we will concentrate
on those processes that are sensitive to short-distance physics and
probe the flavordynamics sector of the standard model.
\\
After this introduction we briefly describe, in section 2, the 
necessary theoretical framework. The decays $K^+\to\pi^+\nu\bar\nu$
and $K_L\to\pi^0\nu\bar\nu$ form the subject of section 3.
Sections 4, 5 and 6 deal with $K_L\to\pi^0e^+e^-$, $K_L\to\mu^+\mu^-$
and $K^+\to\pi^+\mu^+\mu^-$, respectively. A few further topics
(chiral perturbation theory, additional possibilities to test CP violation,
SM forbidden decays) are briefly addressed in section 7.
We conclude with a summary in section 8.

\section{THEORETICAL FRAMEWORK}

\begin{figure}[t]
 \vspace{5cm}
\includegraphics{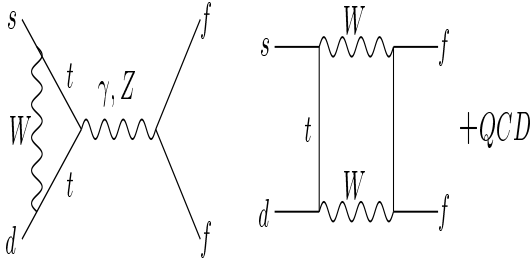}
 \caption{
      Typical diagrams inducing rare K decays in the standard model.
    \label{sdfffig} }
\end{figure}

Loop-induced flavor-changing neutral current (FCNC) processes,
as they appear at higher order in the standard electroweak theory,
can give rise to rare decays of K mesons (Fig. \ref{sdfffig}).
The calculation of decay rates and branching fractions is based on the
construction of low-energy effective Hamiltonians using operator
product expansion (OPE) techniques. This method provides a systematic
approximation to the full SM dynamics and is appropriate if the
typical scale of external momenta is small compared to the mass scale
of heavy virtual particles ($M_W$, $m_t$, $m_c$) in the loop.
For K decays this condition is well satisfied. Schematically
the effective Hamiltonians take the form
\begin{equation}\label{heff}
{\cal H}_{eff}=\frac{G_F}{\sqrt{2}}\sum_i V_{CKM} C_i(M_W,\mu)\cdot Q_i
\end{equation}
where $V_{CKM}$ denotes the appropriate CKM parameters, $C_i$ the
Wilson coefficients and $Q_i$ local four-fermion operators.  
Generally the leading terms in the OPE, represented by four-fermion
operators of lowest dimension (six), are sufficient for all practical
purposes. For the example sketched in Fig. \ref{sdfffig}, the
$Q_i$ typically have the form $(\bar sd)_{V-A}(\bar ff)_{V-A}$.
The analysis of higher-dimensional operators, corresponding to
subleading contributions in the OPE, is discussed in \cite{PIV}.
\\
The result of the formal OPE, eq. (\ref{heff}), has the intuitive
interpretation of an effective, Fermi-type theory with the operators
$Q_i$ playing the role of interaction vertices and the coefficient
functions $C_i$ representing the corresponding coupling constants.
Beyond being just a convenient approximation, the formalism based
on (\ref{heff}) exhibits a very crucial conceptual feature:
It provides a factorization of the full amplitude into a short-distance
part, described by the Wilson coefficients $C_i$, and a long-distance
contribution, contained in the hadronic matrix elements of local
operators $Q_i$. The $C_i$ comprise all the physics at short
distances (above the factorization scale $\mu\approx 1 GeV$),
in particular the dependence on $M_W$, on heavy quark masses
($m_t$, $m_c$) and CKM couplings ($V_{td}$, $V_{ts}$). This part
is calculable in perturbation theory, including the short-distance
QCD effects, which may require renormalization group (RG) improvement
if large logarithms (e.g. $\ln(M_W/m_c)$) are present. All the
non-perturbative long-distance dynamics (below scale $\mu$) is factored
into the matrix elements of $Q_i$ between the given external states
and can be treated separately.
In this manner the OPE helps to disentangle the complicated interplay
of strong and weak interactions in FCNC decays.

Continuous progress in both theory and experiment has lead to an improved
understanding of weak decays of hadrons. Simultaneously important
SM parameters are becoming increasingly better under control. One of
the more recent highlights is certainly the discovery of the top-quark,
which has a crucial impact on the field of rare kaon processes.
The top mass is already quite accurately determined. The pole mass
directly measured in experiment reads $m_{t,pole}=(175\pm 6)GeV$ \cite{TIP},
which translates into a running $\overline{MS}$ mass 
$\bar m_t(m_t)\equiv m_t=(167\pm 6)GeV$. The latter definition is the one
used in FCNC processes, where top appears as a virtual particle.
\\
Further quantities important for analyzing rare K decays are
the CKM matrix elements $V_{cb}$ and $|V_{ub}/V_{cb}|$. From exclusive
\cite{NEU} and inclusive \cite{SUV,BBB} $b\to c$ transitions one
can extract $V_{cb}=0.040\pm 0.003$. The situation with
$|V_{ub}/V_{cb}|$ is more difficult and the uncertainty is still
substantial, $|V_{ub}/V_{cb}|\simeq 0.08\pm 0.02$. However the
discovery of $B\to(\pi,\varrho)l\nu$ decays at CLEO \cite{CLEO}
is encouraging and should eventually allow improvements on this issue.
\\
During recent years progress has also been made in computing the effective
Hamiltonians for weak decays. In leading order of RG improved
perturbation theory the leading logarithmic QCD effects of the form
$(\alpha_s \ln(M_W/\mu))^n$ are taken into account to all orders
$n=0, 1, 2,\ldots$ in evaluating the Wilson coefficients $C_i$.
This resummation is necessary due to the presence of large
logarithms $\ln(M_W/\mu)$, compensating the smallness of $\alpha_s$,
and these corrections are counted as terms of ${\cal O}(1)$. At the
next-to-leading order (NLO) the corrections of relative order
${\cal O}(\alpha_s)$ ($\alpha_s(\alpha_s \ln(M_W/\mu))^n$) are
consistently included. These NLO calculations are by now available
for essentially all important rare and CP violating processes.
\\
Several quantitative as well as conceptual improvements are brought
about by a full NLO analysis.
First, going beyond the leading order result and including the
relative ${\cal O}(\alpha_s)$ correction is necessary to assess
the validity of the perturbative approach. Furthermore, without a
NLO calculation a meaningful use of the scheme-specific QCD scale
$\Lambda_{\overline{MS}}$ is not possible, since the distinction
between various schemes shows up only at NLO. Likewise the ambiguity
related to unphysical renormalization scale ($\mu$) dependences,
a theoretical error owing to the truncation of the perturbation series,
can be reduced by including higher order corrections.
In some cases the phenomenologically interesting top-quark mass dependence
is altogether a NLO effect and thus requires the full NLO analysis to
be included in an entirely satisfactory way. This is the case, for example,
with $K_L\to\pi^0e^+e^-$. In other decays, like $K\to\pi\nu\bar\nu$,
the $m_t$-dependence appears already at leading order. In this situation
the NLO calculation is necessary for a meaningful distinction between
various possible definitions of the top-quark mass, which differ at
${\cal O}(\alpha_s)$. As we have seen above, the difference between
the running mass $m_t(m_t)$ and the pole mass $m_{t,pole}$, for instance,
is about $8 GeV$, a sizable amount that already exceeds the experimental
error $\delta m_t\approx 6 GeV$.

The subject of next-to-leading order QCD corrections to weak decays
has been  reviewed in detail in \cite{BBL}. More information on this
topic and references may be found in this article.

\section{THE RARE DECAYS $K^+\to\pi^+\nu\bar\nu$ AND
$K_L\to\pi^0\nu\bar\nu$}

\begin{figure}[t]
 \vspace{5cm}
\includegraphics{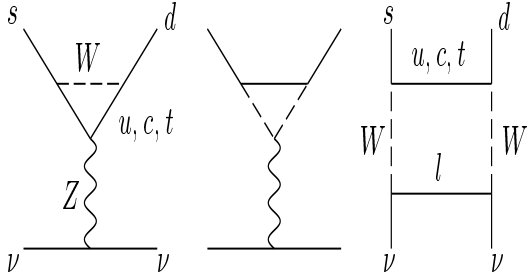}
 \caption{
      Leading order electroweak diagrams contributing to
      $K\to\pi\nu\bar\nu$ in the standard model.
    \label{kpnnfig} }
\end{figure}

The decays $K\to\pi\nu\bar\nu$ proceed through flavor changing
neutral current effects. These arise in the standard model only at
second (one-loop) order in the electroweak interaction
(Z-penguin and W-box diagrams, Fig. \ref{kpnnfig}) 
and are additionally GIM suppressed.
The branching fractions are thus very small, at the level of
$10^{-10}$, which makes these modes rather challenging to detect.
However, $K\to\pi\nu\bar\nu$ have long been known to be reliably
calculable, in contrast to most other decay modes of interest.
A measurement of $K^+\to\pi^+\nu\bar\nu$ and $K_L\to\pi^0\nu\bar\nu$
will therefore be an extremely useful test of flavor physics.
Over the recent years important refinements have been added to the
theoretical treatment of $K\to\pi\nu\bar\nu$. These have helped to
precisely quantify the meaning of the term `clean' in this context
and have reinforced the unique potential of these observables.
Let us briefly summarize the main aspects of why $K\to\pi\nu\bar\nu$
is theoretically so favorable and what recent developments have
contributed to emphasize this point.
\begin{itemize}
\item
First, $K\to\pi\nu\bar\nu$ is a semileptonic decay. 
The relevant hadronic
matrix elements $\langle\pi|(\bar sd)_{V-A}|K\rangle$ are just
matrix elements of a current operator between hadronic states, which are
already considerably simpler objects than the matrix elements of
four-quark operators encountered in many other observables
($K-\bar K$ mixing, $\varepsilon'/\varepsilon$).
Moreover, they are related to the matrix element
\begin{equation}\label{sume}
\langle\pi^0|(\bar su)_{V-A}|K^+\rangle
\end{equation} 
by isospin symmetry.
The latter quantity can be extracted from the well measured leading
semileptonic decay $K^+\to\pi^0 l\nu$. Although isospin is a fairly
good symmetry, it is still broken by the small 
difference between the masses of up- and down-quarks and by 
electromagnetism. 
These sources of isospin breaking manifest themselves
in differences of the neutral versus charged kaon (and pion) masses
(affecting phase space), corrections to the isospin limit in the
formfactors and electromagnetic radiative effects.
Marciano and Parsa \cite{MP} have analyzed these corrections and
found an overall reduction in the branching ratio by $10\%$ for
$K^+\to\pi^+\nu\bar\nu$ and by $5.6\%$ for $K_L\to\pi^0\nu\bar\nu$.
\item
Long-distance contributions are systematically suppressed 
as ${\cal O}(\Lambda^2_{QCD}/m^2_c)$ compared to the 
charm contribution (which is part of the short-distance amplitude).
This feature is related to the hard ($\sim m^2_c$) GIM suppression
pattern exhibited by the Z-penguin and W-box diagrams, and the absence
of virtual photon amplitudes. Long-distance contributions have
been examined quantitatively \cite{RS,HL,LW,GHL,FAJ} and shown to be
indeed negligible numerically 
(below $\approx 5\%$ of the charm amplitude).
\item
The preceeding discussion implies that $K\to\pi\nu\bar\nu$ are
short-distance dominated (by top- and charm-loops in general).
The relevant short-distance QCD effects can be treated in
perturbation theory and have been calculated at next-to-leading
order \cite{BB2,BB3}. This allowed to substantially reduce
(for $K^+$) or even practically eliminate ($K_L$) the leading
order scale ambiguities, which are the dominant uncertainties in
the leading order result.
\end{itemize}

In Table \ref{kpnntab} we have summarized some of the main
features of $K^+\to\pi^+\nu\bar\nu$ and $K_L\to\pi^0\nu\bar\nu$.
\begin{table*}[hbt]
\setlength{\tabcolsep}{1.5pc}
\newlength{\digitwidth} \settowidth{\digitwidth}{\rm 0}
\catcode`?=\active \def?{\kern\digitwidth}
\caption{Compilation of important properties and results for
$K\to\pi\nu\bar\nu$}
\label{kpnntab}
\begin{tabular*}{\textwidth}{@{}l@{\extracolsep{\fill}}rrrr}
\hline
                & $K^+\to\pi^+\nu\bar\nu$ 
                 & $K_L\to\pi^0\nu\bar\nu$ \\
\hline
& CP conserving & CP violating \\
CKM & $V_{td}$ & $\mbox{Im} V^*_{ts}V_{td}\sim J_{CP}\sim\eta$ \\
contributions & top and charm & only top \\
scale uncert. (BR) & $\pm 20\%$ (LO) $\to \pm 5\%$ (NLO) &
                     $\pm 10\%$ (LO) $\to \pm 1\%$ (NLO) \\
BR (SM) & $(0.9\pm 0.3)\cdot 10^{-10}$&$(2.8\pm 1.7)\cdot 10^{-11}$ \\
exp. limit & $< 2.4\cdot 10^{-9}$ BNL 787 \cite{ADL}
           & $< 5.8\cdot 10^{-5}$ FNAL 799 \cite{WEA} \\
\hline
\end{tabular*}
\end{table*}

The neutral mode proceeds through CP violation in the standard model.
This is due to the definite CP properties of $K^0$, $\pi^0$ and
the hadronic transition current $(\bar sd)_{V-A}$. The violation of
CP symmetry in $K_L\to\pi^0\nu\bar\nu$ arises through interference
between $K^0-\bar K^0$ mixing and the decay amplitude. This mechanism
is sometimes refered to as mixing-induced CP violation. 
By itself, it could a priori be attributed to a superweak interaction.
However, any difference in the magnitude of this mixing-induced
CP violation between two different $K_L$ decay modes can not, and is
therefore a signal of direct CP violation. Now, the effect of
mixing-induced CP violation is already known for e.g. $K_L\to\pi^+\pi^-$.
In this case it can be measured by 
$\eta_{+-}=A(K_L\to\pi^+\pi^-)/A(K_S\to\pi^+\pi^-)\approx\varepsilon_K$,
which is of the order ${\cal O}(10^{-3})$. By contrast,
$\eta_{\pi^0\nu\bar\nu}=A(K_L\to\pi^0\nu\bar\nu)/A(K_S\to\pi^0\nu\bar\nu)$
is of ${\cal O}(1)$ in the SM, far larger than $\eta_{+-}$.  
Thus the standard model decay $K_L\to\pi^0\nu\bar\nu$ is
a signal of almost pure direct CP violation, revealing an effect
that can not be attributed to CP violation in the $K-\bar K$
mass matrix alone (in which case $\eta_{+-}=\eta_{\pi^0\nu\bar\nu}$).
\\
While already $K^+\to\pi^+\nu\bar\nu$ can be reliably calculated,
the situation is even better for $K_L\to\pi^0\nu\bar\nu$. Since
only the imaginary part of the amplitude (in standard phase
conventions) contributes, the charm sector, in $K^+\to\pi^+\nu\bar\nu$
the dominant source of uncertainty, is completely negligible for
$K_L\to\pi^0\nu\bar\nu$ ($0.1\%$ effect on the branching ratio).
Long-distance contributions 
($\;\raisebox{-.4ex}{\rlap{$\sim$}} \raisebox{.4ex}{$<$}\; 0.1\%$)  
and also the indirect CP violation effect  
($\;\raisebox{-.4ex}{\rlap{$\sim$}} \raisebox{.4ex}{$<$}\; 1\%$)  
are likewise negligible. In summary, the total theoretical
uncertainties, from perturbation theory in the top sector
and in the isospin breaking corrections, are safely below
$2-3\%$ for $B(K_L\to\pi^0\nu\bar\nu)$. This makes this decay
mode truly unique and very promising for phenomenological
applications. (Note that the range given as the standard model
prediction in Table \ref{kpnntab} arises from our, at present,
limited knowledge of standard model parameters (CKM), and not
from intrinsic uncertainties in calculating $B(K_L\to\pi^0\nu\bar\nu)$).

\begin{figure}[t]
 \vspace{5cm}
\includegraphics{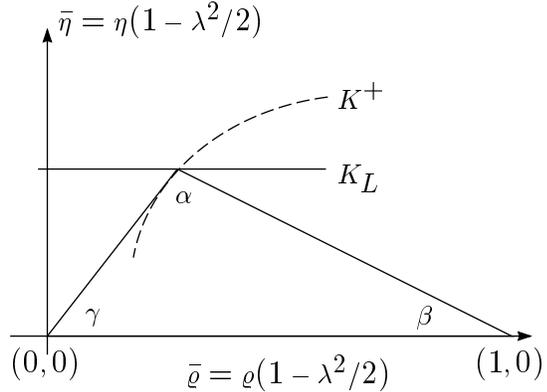}
 \caption{
      Unitarity triangle from $K\to\pi\nu\bar\nu$.
    \label{utkpnfig} }
\end{figure}

With a measurement of $B(K^+\to\pi^+\nu\bar\nu)$ and 
$B(K_L\to\pi^0\nu\bar\nu)$ available, very interesting phenomenological
studies could be performed. 
For instance, $B(K^+\to\pi^+\nu\bar\nu)$ and 
$B(K_L\to\pi^0\nu\bar\nu)$ together determine the unitarity triangle
(Wolfenstein parameters $\varrho$ and $\eta$) 
completely (Fig. \ref{utkpnfig}). 
The expected accuracy with $\pm 10\%$ branching ratio measurements is
comparable to the one that can be achieved by CP violation studies
at $B$ factories before the LHC era \cite{BB96}.
The quantity $B(K_L\to\pi^0\nu\bar\nu)$ by itself offers probably the
best precision in determining $\mbox{Im} V^*_{ts}V_{td}$ or,
equivalently, the Jarlskog parameter
\begin{equation}\label{jcp}
J_{CP}=\mbox{Im}(V^*_{ts}V_{td}V_{us}V^*_{ud})=
\lambda\left(1-\frac{\lambda^2}{2}\right)\mbox{Im}\lambda_t
\end{equation}
The prospects here are even better than for $B$ physics at the LHC.
As an example, let us assume the following results will be
available from B physics experiments
\begin{eqnarray}\label{lhcb}
\sin 2\alpha &=& 0.40\pm 0.04 \nonumber \\ 
\sin 2\beta &=& 0.70\pm 0.02 \\
V_{cb} &=& 0.040\pm 0.002 \nonumber
\end{eqnarray}
The small errors quoted for $\sin 2\alpha$ and $\sin 2\beta$ from
CP violation in $B$ decays require precision measurements at the LHC.
In the case of $\sin 2\alpha$ we have to assume in addition that the
theoretical problem of `penguin-contamination' can be resolved.
These results would then imply
$\mbox{Im}\lambda_t=(1.37\pm 0.14)\cdot 10^{-4}$.
On the other hand, a $\pm 10\%$ measurement 
$B(K_L\to\pi^0\nu\bar\nu)=(3.0\pm 0.3)\cdot 10^{-11}$ together with
$m_t(m_t)=(170\pm 3) GeV$ would give
$\mbox{Im}\lambda_t=(1.37\pm 0.07)\cdot 10^{-4}$. If we are optimistic
and take $B(K_L\to\pi^0\nu\bar\nu)=(3.0\pm 0.15)\cdot 10^{-11}$,
$m_t(m_t)=(170\pm 1) GeV$, we get
$\mbox{Im}\lambda_t=(1.37\pm 0.04)\cdot 10^{-4}$, a truly remarkable
accuracy. The prospects for precision tests of the standard model
flavor sector will be correspondingly good.

The charged mode $K^+\to\pi^+\nu\bar\nu$ is being currently
pursued by Brookhaven experiment E787. The latest published result
\cite{ADL} gives an upper limit which is about a factor 25 above
the standard model range. Several improvements have been implemented
since then and the SM sensitivity is expected to be reached in the
near future \cite{AGS2}. For details see also \cite{MEY}.
Recently an experiment has been proposed to measure 
$K^+\to\pi^+\nu\bar\nu$ at the Fermilab Main Injector \cite{CCTR}.
Concerning $K_L\to\pi^0\nu\bar\nu$, a proposal exists at
Brookhaven (BNL E926) to measure this decay at the AGS 
with a sensitivity of ${\cal O}(10^{-12})$ \cite{AGS2}.
There are furthermore plans to pursue this mode with comparable
sensitivity at Fermilab \cite{KAMI} and KEK \cite{ISS,SAS}.
It will be very exciting to follow the development and outcome of
these ambitious projects.

\section{$K_L\to\pi^0e^+e^-$}

This decay mode has obvious similarities with $K_L\to\pi^0\nu\bar\nu$
and the apparent experimental advantage of charged leptons, rather than
neutrinos, in the final state. However there are a number of quite
serious difficulties associated with this very fact. Unlike neutrinos the
electron couples to photons. As a consequence the amplitude, which 
was essentially purely short-distance in $K_L\to\pi^0\nu\bar\nu$, becomes
sensitive to poorly calculable long-distance physics (photon penguin).
Simultaneously the importance of indirect CP violation 
($\sim\varepsilon$) is strongly enhanced and furthermore a long-distance
dominated, CP conserving amplitude with two-photon intermediate state
can contribute significantly. Treating $K_L\to\pi^0e^+e^-$
theoretically one is thus faced with the need to disentangle three
different contributions of roughly the same order of magnitude.
\begin{itemize}
\item
Direct CP violation: This part is short-distance in character,
theoretically clean and has been analyzed at next-to-leading order
in QCD \cite{BLMM}. Taken by itself this mechanism leads 
within the standard model to a
$K_L\to\pi^0e^+e^-$ branching ratio of $(4.5\pm 2.6)\cdot 10^{-12}$.
\item
Indirect CP violation: This amplitude is determined through
$\sim\varepsilon\cdot A(K_S\to\pi^0e^+e^-)$. The $K_S$ amplitude is
dominated by long-distance physics and has been investigated in
chiral perturbation theory \cite{EPR,BRP,DG}. Due to unknown counterterms
that enter this analysis a reliable prediction is not possible at
present. The situation would improve with a measurement of
$B(K_S\to\pi^0e^+e^-)$, which could become possible at DA$\Phi$NE.
Present day estimates for $B(K_L\to\pi^0e^+e^-)$ due to indirect
CP violation alone give typically values of
$(1-5)\cdot 10^{-12}$.
\item
The CP conserving two-photon contribution is again long-distance
dominated. It has been analyzed by various authors \cite{DG,CEP,HS}.
The estimates are typically a few $10^{-12}$. Improvements in this sector
might be possible by further studying the related decay
$K_L\to\pi^0\gamma\gamma$ whose branching ratio has already been
measured to be $(1.7\pm 0.3)\cdot 10^{-6}$. 
\end{itemize}

Originally it had been hoped for that the direct CP violating
contribution is dominant. Unfortunately this could so far not be
unambiguously established and requires further study.
\\
Besides the theoretical problems, $K_L\to\pi^0e^+e^-$ is also very hard
from an experimental point of view. The expected branching ratio 
is even smaller than for $K_L\to\pi^0\nu\bar\nu$. Furthermore a
serious irreducible physics background from the radiative mode
$K_L\to e^+e^-\gamma\gamma$ has been identified, which poses additional
difficulties \cite{LV}. A background subtraction seems necessary,
which is possible with enough events. 
Additional information could in principle also be gained by studying
the electron energy asymmetry \cite{DG,HS} or the time evolution
\cite{DG,LIT,KP}.

\section{$K_L\to\mu^+\mu^-$}

\begin{figure}[t]
 \vspace{5cm}
\includegraphics{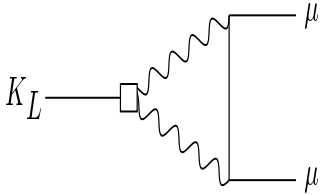}
 \caption{
     Long-distance contribution to $K_L\to\mu^+\mu^-$ from
     the two-photon intermediate state.
    \label{klmmfig} }
\end{figure}

$K_L\to\mu^+\mu^-$ receives a short-distance contribution from
Z-penguin and W-box graphs similarly to $K\to\pi\nu\bar\nu$. This
part of the amplitude is sensitive to the Wolfenstein parameter
$\varrho$. In addition $K_L\to\mu^+\mu^-$ proceeds through a
long-distance contribution with two-photon intermediate state,
which actually dominates the decay completely (Fig. \ref{klmmfig}).
The long-distance
amplitude consists of a dispersive ($A_{dis}$) and an absorbtive
contribution ($A_{abs}$). The branching fraction can thus be written
\begin{equation}\label{bklma}
B(K_L\to\mu^+\mu^-)=|A_{SD}+A_{dis}|^2 + |A_{abs}|^2
\end{equation}
Using $B(K_L\to\gamma\gamma)$ it is possible to extract
$|A_{abs}|^2=(6.8\pm 0.3)\cdot 10^{-9}$ \cite{LV}.
$A_{dis}$ on the other hand can not be calculated accurately at 
present and the estimates are strongly model dependent 
\cite{BMS,GN,BEG,KO,EKP}.
This is rather unfortunate, in particular since 
$B(K_L\to\mu^+\mu^-)$, unlike most other rare decays, has already
been measured, and this with very good precision
\begin{eqnarray}\label{bklmex}
\lefteqn{B(K_L\to\mu^+\mu^-)=} \\
 & & \left\{ \begin{array}{ll}
        (6.9\pm 0.4)\cdot 10^{-9} & \mbox{BNL 791 \cite{HEIN}}\\
        (7.9\pm 0.7)\cdot 10^{-9} & \mbox{KEK 137 \cite{AKA}}
        \end{array} \right. \nonumber
\end{eqnarray}
For comparison we note that 
$B(K_L\to\mu^+\mu^-)_{SD}=(1.3\pm 0.6)\cdot 10^{-9}$ is the
expected branching ratio in the standard model based on the
short-distance contribution alone. Due to the fact that $A_{dis}$
is largely unknown, $K_L\to\mu^+\mu^-$ is at present not a very
useful constraint on CKM parameters.
Some improvement of the situation might be expected from measuring
the decay $K_L\to\mu^+\mu^-e^+e^-$, which could lead to a better
understanding of the $K_L\to\gamma^*\gamma^*$ vertex.
First results obtained at Fermilab (E799) give
$B(K_L\to\mu^+\mu^-e^+e^-)=(2.9^{+6.7}_{-2.4})\cdot 10^{-9}$.

\section{$K^+\to\pi^+\mu^+\mu^-$}

The rare decay $K^+\to\pi^+\mu^+\mu^-$ has recently been measured
at Brookhaven (BNL 787)  with a branching ratio \cite{B787}
\begin{equation}\label{bkpmm}
B(K^+\to\pi^+\mu^+\mu^-)=(5.0\pm 0.4\pm 0.6)\cdot 10^{-8}
\end{equation}
This compares well with the estimate from chiral perturbation theory
$B(K^+\to\pi^+\mu^+\mu^-)=(6.2^{+0.8}_{-0.6})\cdot 10^{-8}$ \cite{PIC}.
The branching ratio is completely determined by the long-distance 
contribution arising from the one-photon exchange amplitude.
A short-distance amplitude from Z-penguin and W-box diagrams (similar to
Fig. \ref{sdfffig}) also exists, but is smaller than the long-distance
part by three orders of magnitude and does therefore not play any role
in the total rate. However, while the muon pair couples via a vector
current $(\bar\mu\mu)_V$ in the photon amplitude, the electroweak
short-distance mechanism also contains an axial vector piece 
$(\bar\mu\mu)_A$. The interference term between these contributions
is odd under parity and gives rise to a parity-violating longitudinal
$\mu^+$ polarization, which can be observed as an asymmetry
$\Delta_{LR}=(\Gamma_R-\Gamma_L)/(\Gamma_R+\Gamma_L)$
\cite{SW,LWS,BGT,BB5}.
$\Gamma_{R(L)}$ denotes the rate of producing a right- (left-) handed
$\mu^+$ in $K^+\to\pi^+\mu^+\mu^-$ decay. The effect occurs for
a $\mu^-$ instead of $\mu^+$ as well, but the polarization measurement
is much harder in this case.
\\
$\Delta_{LR}$ is sensitive to the Wolfenstein parameter $\varrho$.
It is a cleaner observable than $K_L\to\mu^+\mu^-$, although some
contamination through long-distance contributions can not be excluded
\cite{LWS}. At any rate, $\Delta_{LR}$ will be an interesting observable
to study if a sensitive polarization measurement becomes feasible.
The standard model expectation is typically around
$\Delta_{LR}\sim 0.5\%$.

\section{FURTHER TOPICS}

\subsection{Tests of chiral perturbation theory}
 
Chiral perturbation theory provides a systematic framework to treat
the strong interactions of kaons and pions at low energies. 
Long-distance dominated rare and radiative kaon decays, such as
$K_S\to\gamma\gamma$, $K_L\to\pi^0\gamma\gamma$, $K^+\to\pi^+\gamma\gamma$
or $K^+\to\pi^+l^+l^-$, offer ample testing ground for this approach.
These studies are interesting and important in their own right and may
also be helpful for the extraction of short-distance effects. An
example of the latter case is the muon polarization asymmetry
$\Delta_{LR}$ in $K^+\to\pi^+\mu^+\mu^-$ discussed in the previous
section.
\\
The application of chiral perturbation theory to rare K decays has been
most recently reviewed in \cite{PIC,ECK}, where further references 
can be found. Other useful accounts of this subject, including experimental
tests are given in \cite{LV,RW}.

\subsection{CP Violation}

In addition to the more commonly discussed  observables of CP violation
in K decays, like $\varepsilon$ \cite{ROS},
$\varepsilon'/\varepsilon$ \cite{CIU} or
$K_L\to\pi^0\nu\bar\nu$, other options for probing the
CP asymmetry of nature in rare kaon processes have been proposed in the
literature. For instance, any difference between the decay rates of
$K^+\to\pi^+\gamma\gamma$ and $K^-\to\pi^-\gamma\gamma$ would signal
direct CP violation. The same applies to
$K^+\to\pi^+e^+e^-$ and $K^-\to\pi^-e^+e^-$. Although a theoretical
treatment of these long-distance dominated modes is not easy, observation
of a clear CP violating effect would certainly be very interesting.
\\
Another example is the longitudinal $\mu^+$ polarization in
$K_L\to\mu^+\mu^-$, which violates CP. The standard model prediction
is rather reliable in this case and one expects 
$(\Gamma_R-\Gamma_L)/(\Gamma_R+\Gamma_L)\approx 2\cdot 10^{-3}$
\cite{EP}.
\\
More information on these topics may be found in the review articles
\cite{LV,RW,WW}.

\begin{table*}[hbt]
\caption{Summary of present status of selected rare kaon decays.}
\label{sumtab}
\begin{tabular*}{\textwidth}{@{}l@{\extracolsep{\fill}}cccc}
\hline 
& $K_L\to\mu^+\mu^-$ & $K_L\to\pi^0e^+e^-$
& $K^+\to\pi^+\nu\bar\nu$ & $K_L\to\pi^0\nu\bar\nu$ \\
\hline
theoret. status & $- -$ & $+ -$ & $+ +$ & $+ +$ \\
BR (SM) & $(1.3\pm 0.6)\cdot 10^{-9}$  &
$(4.5\pm 2.6)\cdot 10^{-12}$  &
$(0.9\pm 0.3)\cdot 10^{-10}$ &
$(2.8\pm 1.7)\cdot 10^{-11}$ \\
& (SD) & (dir. CPV) & & \\
BR (exp) & $(7.2\pm 0.5)\cdot 10^{-9}$  &
$< 4.3\cdot 10^{-9}$  &
$< 2.4\cdot 10^{-9}$ &
$< 5.8\cdot 10^{-5}$ \\
\hline
\end{tabular*}
\end{table*}

\subsection{SM forbidden decays}

As mentioned in the introduction, lepton flavor violating decays of
kaons can serve as sensitive, albeit indirect, probes of very
high energy scales \cite{LV,RW}. 
The current situation may be characterized by the
following limits on branching ratios that have been established in
experiments at Brookhaven and Fermilab \cite{PDG}:
\begin{eqnarray}\label{kmue}
B(K_L\to\mu e) &<& 3.3\cdot 10^{-11} \quad\mbox{BNL 791} \nonumber \\ 
B(K^+\to\pi^+\mu^+ e^-) &<& 2.1\cdot 10^{-10}  
\quad\mbox{BNL 777}\nonumber \\ 
B(K_L\to\pi^0\mu e) &<& 3.2\cdot 10^{-9} \quad\mbox{FNAL 799}\nonumber \\
\end{eqnarray}
Forbidden in the standard model, those decays could be induced in
extensions that violate lepton flavor. The above branching ratios would
then behave typically as $BR\sim 1/M^4_X$, where $M_X$ is the mass of
an exotic heavy boson mediating the interaction at tree level.
Improvements of the experiments mentioned before aim to reach a sensitivity
in the branching ratios of $\sim 10^{-12}$. This corresponds to
typically $M_X\sim 100 TeV$, which, although somewhat model dependent,
is still quite an impressive figure. The window on such extremely short
distances thus provided makes this class of decays certainly worth
pursuing.

\section{SUMMARY}

The field of rare kaon decays offers a broad range of interesting
topics, ranging from chiral perturbation theory, over standard model
flavor dynamics to exotic phenomena, thereby covering scales from
$\Lambda_{QCD}$ to the weak scale ($M_W$, $m_t$) and beyond to maybe
several hundred $TeV$. In the present talk we have focussed on the
flavor physics of the standard model and those processes that can be
used to test it. Several promising examples of short-distance sensitive 
decay modes exist, whose experimental study will provide important clues
on flavordynamics.
On the theoretical side, progress has been achieved over recent years
in the calculation of effective Hamiltonians, which by now include
the complete NLO QCD effects in essentially all cases of practical
interest.
The current status of four particularly important decay modes,
$K_L\to\mu^+\mu^-$, $K_L\to\pi^0e^+e^-$, $K^+\to\pi^+\nu\bar\nu$
and $K_L\to\pi^0\nu\bar\nu$, is briefly summarized in Table
\ref{sumtab}.
The SM predictions for the branching ratios are determined by the
usual analysis that fits the CKM phase $\delta$ from the experimental
value of the kaon CP violation parameter $\varepsilon$, and requires
$m_t$, $V_{cb}$, $|V_{ub}/V_{cb}|$ and the kaon bag parameter $B_K$
as main input.  
The SM numbers in Table \ref{sumtab} are from ref. \cite{BJLnew} and use 
$\bar m_t(m_t)=(167\pm 6)GeV$, $V_{cb}=0.040\pm 0.003$, 
$|V_{ub}/V_{cb}|= 0.08\pm 0.02$ and (for the RG invariant
bag parameter) $B_K=0.75\pm 0.15$.
For $K_L\to\mu^+\mu^-$ the branching ratio prediction refers only
to the short-distance part, for $K_L\to\pi^0e^+e^-$ only to the
contribution from direct CP violation.
\\
The decay $K_L\to\mu^+\mu^-$ is already measured quite accurately;
unfortunately a quantitative use of this result for the determination
of CKM parameters is strongly limited by large hadronic uncertainties.
For $K_L\to\pi^0e^+e^-$ there are some theoretical and experimental
difficulties, but improvements might be possible.
The situation looks very bright for $K\to\pi\nu\bar\nu$. The charged
mode is experimentally already `around the corner' and its very
clean status promises useful results on $V_{td}$. Finally the decay
$K_L\to\pi^0\nu\bar\nu$ is a particular highlight in this field.
It could serve e.g. as the ideal measure of the Jarlskog parameter
$J_{CP}$. Measuring the branching ratio is a real experimental
challenge, but definitely worth the effort.

It is to be expected that rare kaon decay phenomena will in the
future continue to contribute substantially to our understanding
of the fundamental interactions and it is quite conceivable that
exciting surprizes await us along the way.

\end{document}